\documentstyle[prl,aps,epsf,twocolumn,eqsecnum]{revtex}
\begin{document}
\draft
\title{Coherent nuclear motion in a condensed-phase environment:
Wave-packet approach and pump-probe spectroscopy}
\author{Lothar M\"uhlbacher, Andreas Lucke, and Reinhold Egger}
\address{Fakult\"at f\"ur Physik, Albert-Ludwigs-Universit\"at,  
D-79104 Freiburg, Germany}
\date{Date: \today}
\maketitle
\begin{abstract}
A quantum-mechanical Gaussian wave-packet approach to the theoretical
description of nuclear motions in a condensed-phase environment is developed.
General expressions for the time-dependent reduced density matrix are given for
a harmonic potential surface, and the exact quantum dynamics is found for a
microscopic system-plus-bath model. Particular attention is devoted to the
influence of initial correlations between system and bath for the outcome of a
pump-probe experiment.  We show that the standard factorized preparation,
compared to a more realistic correlated preparation, leads to significantly
different stimulated emission spectra at high temperatures.
Recent experiments for the reaction center are analyzed using this
formalism.
\end{abstract}

\narrowtext
\section{Introduction}

The notion of wave packets is intimately connected with the foundations of
quantum mechanics itself.  Despite of their importance during the initial
stages of the development of quantum theory \cite{schr}, this concept has been
quickly overturned by the powerful and elegant operator formalism. Only during
the past decade interest in wave packets has emerged again, primarily triggered
by the widespread availability of ultrafast femtosecond laser pulse techniques
\cite{beddard}.  Wave packets are by today a standard tool employed to explain
many different features in chemistry and physics, e.g., chemical reaction
dynamics \cite{wp,domcke,mukamel}, oscillatory motion of a coherent
Bose-Einstein condensate \cite{bose}, highly-excited Rydberg states in atoms
\cite{alber}, or electron-hole excitations in semiconductors
\cite{solid,weiss}.  The concept of wave packets is sometimes applied to
rationalize experimental data even though the reaction occurs under
condensed-phase conditions, where the relevant reaction coordinate (e.g.,
describing the nuclear motion) of the wave packet (``system'') may be strongly
coupled to other modes (``environment'' or ``bath'').  At this point one may
ask whether it makes sense to use a wave-packet description for the reaction
coordinate dynamics even if strong coupling to solvent modes is present.
First, the description of the wave packet in terms of a wavefunction is moot,
and one has to use the reduced density matrix.  Second, the bath leads to a
damping of the wave packet and could cause the complete loss of coherence. In
that case, the use of wave packets would be rather restricted.  It is one of
the purposes of this paper to clarify to what extent the wave-packet concept is
applicable in the presence of strong system-bath coupling.

The dissipation acting on the wave packet can have many different microscopic
origins. In gas-phase reactions, one typically has rather weak damping due to
coupling to the vacuum modes of the electromagnetic field (spontaneous
emission), or due to collisions with other molecules.  In contrast, dissipation
can become decisive in condensed-phase reactions where one has strong coupling
of the system
to solvent or protein polarization modes.  Considering previous theoretical
treatments of gas-phase reactions, dissipation was mostly ignored or at best
incorporated within the framework of the Bloch equations \cite{bloch} or the
Redfield equations \cite{redfield}, where the latter allow to retain memory
effects.  However, as such an approach relies on perturbation
theory in the system-bath coupling, its application to condensed-phase
reactions characterized by strong system-bath coupling remains questionable. 
Other methods are based on classical molecular dynamics (MD) simulations 
\cite{so} or projection operator techniques \cite{mukamel}.

In general, the problem of dissipative wave-packet motion is theoretically
quite demanding.  In this paper, we treat a simple model introduced in Sections
\ref{gauss} and \ref{theory} but put particular emphasis on the effects due to
different initial states of the system arising in pump-probe 
spectroscopy experiments. This issue is
shown to be important for a correct description of experimental data on systems
in a condensed-phase environment.  More specifically, one might be tempted to
assume a certain initial preparation which is named ``factorized preparation''
henceforth.  Under the factorized preparation, the density matrix at $t=0$
factorizes into a system part describing the wave packet, and a part
corresponding to the solvent modes.  The wave packet at $t=0$ 
could then correspond
to a pure state, e.g., a Gaussian wave packet.  Here we provide a comparison of
the factorized preparation to the more realistic ``correlated preparation''
which takes into account initial system-bath correlations \cite{grabert88}.  We
stress that such preparation effects cannot be captured by standard
``dissipative wave-packet'' approaches \cite{wp}, which implicitely use the
factorized preparation.

The structure of this paper is as follows.  After presenting
general expressions for  Gaussian wave-packet dynamics in
Sec.~\ref{gauss}, the connection to microscopic system-plus-environment
models is established in Sec.~\ref{theory}.   
In Sec.~\ref{vossec}, we then consider a pump-probe spectroscopy experiment 
involving two harmonic surfaces.  As a practical example,
we analyze the recent stimulated
emission experiments of the bacterial photosynthetic reaction center by
Vos {\em et al.}  \cite{vos93,vos94,vos98}, albeit the theory 
is more generally applicable.  In that section, we also show that
the two initial preparations mentioned above cause pronounced
differences in the emission spectra at high temperatures.
Finally, some conclusions are offered in Sec.~\ref{conc}.
Technical details have been deferred to an appendix.

\section{Gaussian density matrices}\label{gauss}

Let us start with general properties of the time evolution of a Gaussian
reduced density matrix $\rho(t)$.  The Gaussian property implies that the
underlying Hamiltonian is at most quadratic in the system coordinate ($q$) and
momentum ($p$), and imposes certain restrictions for the system-bath coupling.
However, there is neither need to specify a Hamiltonian nor initial conditions
for the system-bath complex at this stage [except
consistency with the Gaussian form of $\rho(t)$]. 
 The spatial representation of the density matrix reads
\begin{equation}\label{rhodef}
\rho(q,q',t)= N^{-1}(t) \exp[-\Sigma(q,q',t)]\;,
\end{equation}
where $\Sigma(q,q',t)$ denotes a quadratic form,
\begin{eqnarray}\label{sigma}
\Sigma(q,q',t) &=& \frac{a_1}{4} q^2 + \frac{a_1^*}{4} q^{\prime 2}
- a_2 q - a_2^* q'
 + \frac{a_3}{2} qq' \\ \nonumber
&+& \frac{[a_2+a_2^*]^2}{a_1+a_1^*+2a_3} \;,
\end{eqnarray}
with arbitrary time-dependent coefficients $a_i(t)$. Furthermore,
the normalization Tr$[\rho]= 1$ is ensured by
choosing 
\[
N(t) = \sqrt{4\pi/[a_1+a_1^*+2a_3]} \;.
\]
While $a_1=a_1^\prime+ia_1^{\prime\prime}$ and 
$a_2=a_2^\prime+ia_2^{\prime\prime}$ can be complex-valued, 
$\rho=\rho^\dagger$ implies a real-valued coefficient $a_3$. 
Therefore we have five independent real-valued
functions, and accordingly there are only five independent expectation
values,
\begin{eqnarray*}
\langle q\rangle &= & 2a_2^\prime/[a_1^\prime+a_3]\;, \\
\langle p \rangle &= & \hbar\left(
a_2^{\prime\prime}
- \frac{a_1^{\prime\prime} a_2^\prime}{a_1^\prime+a_3}
 \right) \;, \\
\langle [\Delta q]^2 \rangle &= & 
\frac{1}{a_1^\prime+a_3} \;, \\
\langle [\Delta p]^2 \rangle
 &= & \frac{\hbar^2}{4} \left(
a_1^\prime - a_3 +\frac{a_1^{\prime\prime 2}}{a_1^\prime+a_3}
\right) \;, \\
\langle [\Delta q, \Delta p]_+ \rangle
 &= & - \hbar\frac{a_1^{\prime\prime}}{a_1^\prime+a_3} \;,
\end{eqnarray*}
where $\Delta q= q-\langle q\rangle$, $\Delta p = p - 
\langle p \rangle$, and $[A,B]_+=AB+BA$.

Let us now assume that the Ehrenfest theorem holds.
This implies 
\begin{eqnarray} \label{ehrenfest}
\frac{d}{dt} \langle q\rangle &=&
\langle p \rangle /m \;,
\\ \frac{d}{dt}
\langle [\Delta q ]^2 \rangle &=&
\langle [\Delta q, \Delta p]_+  \rangle/m\;,
\end{eqnarray}
with the mass $m$.  These equations eliminate
two of the five degrees of freedom. Therefore 
we keep only $\langle q(t) \rangle$,
the variance $\sigma(t)= 2 \langle [\Delta q]^2 (t)\rangle$,
and the quantity 
\begin{equation}\label{theta}
\Theta(t) = \frac{2\hbar}{\sqrt{
8 \sigma\langle[\Delta p]^2\rangle
- m^2 \dot{\sigma}^2}} 
\end{equation}
as independent functions, where $\dot{\sigma}
=d\sigma/dt$.  The functions $a_i(t)$ can be
expressed in terms of these three quantities,
and Eq.~(\ref{sigma}) then takes the form
\begin{eqnarray}\label{sigmare}
&& {\rm Re}\, \Sigma(\langle q \rangle+Q, \langle q\rangle+Q',t) = \\ \nonumber
&&  \frac{1}{2\Theta^2 \sigma} \left[
\frac{1+\Theta^2}{2} (Q^2+Q^{\prime 2}) -
(1-\Theta^2) Q Q^\prime \right] \,, \\ \label{sigmaim}
&&\frac{ \hbar}{m} 
\, {\rm Im}\, \Sigma(\langle q\rangle+Q, \langle q\rangle+Q',t)
= \\ \nonumber
&&  Q^\prime \left(\frac{\dot{\sigma}Q^\prime}{4
\sigma} +  \frac{d\langle q\rangle}{dt} \right)
- Q \left(\frac{\dot{\sigma}Q}{4
\sigma} +  \frac{d\langle q\rangle}{dt} \right) \;.
\end{eqnarray}
The normalization constant
becomes simply $N=\sqrt{\pi \sigma}$.
The real-valued quantity $\Theta$ is always within the
bounds $0<\Theta(t)\leq 1$, with the limiting case
$\Theta=1$ applying to a pure system.
In fact, straightforward algebra yields
\begin{equation}\label{den2}
\Theta(t) = {\rm Tr}\, [\rho^2(t)] \,.
\end{equation}
It is noteworthy that $\Theta(t)$ is in general
an independent quantity, as there is no Ehrenfest
relation expressing $\langle [ \Delta p]^2 (t)\rangle$
solely in terms of $\langle q(t) \rangle$ and
$\sigma(t)$.

By employing the unitary transformation
\begin{equation}\label{unitary}
U(t) = \exp\left[ -\frac{i}{\hbar} m q
\left( \frac{\dot{\sigma} q}{4\sigma} + \frac{d \langle q\rangle}{dt}
\right) 
\right] \exp\left[\frac{i}{\hbar} \langle q\rangle p \right]
\;,
\end{equation}
the density matrix $\widetilde{\rho}(t)=
U\rho U^\dagger$ attains the coordinate-independent
form
\begin{equation}\label{dtr}
\widetilde{\rho}(t) = \widetilde{Z}^{-1} \exp(-\widetilde{\beta}
\widetilde{H} ) \;,
\end{equation}
with the Hamiltonian $\widetilde{H}(t)$
of a harmonic oscillator subject to an effective time-dependent
confinement frequency
\begin{equation}\label{omega}
\widetilde{\Omega}(t) = \frac{\hbar}{m\sigma(t) \Theta(t)} \;.
\end{equation}
The effective inverse temperature $\widetilde{\beta}$ is
\begin{equation}\label{temp}
\widetilde{\beta}(t) = \frac{1}{\hbar\widetilde{\Omega}(t)}
\ln \left(\frac{1+\Theta(t)}{1-\Theta(t)}\right) \;,
\end{equation}
and $\widetilde{Z}(t)= [2 \sinh (\hbar\widetilde{\Omega} \widetilde{\beta}/2)]^{-1}$.
The transformed density matrix (\ref{dtr})
corresponds to the equilibrium density matrix
of the harmonic oscillator $\widetilde{H}$
considered at fixed time $t$.

With the aid of this unitary transformation, it becomes
 easy to find the spectral decomposition of the density
matrix.  Transforming the result back to the original
picture, we obtain
\begin{equation}\label{spectral}
\rho(t) = \sum_{n=0}^\infty \lambda_n(t) |\varphi_n(t)
\rangle\langle\varphi_n(t)| \;,
\end{equation}
where the eigenvalues are given by
\begin{equation}\label{eigen}
\lambda_n(t) = \frac{2\Theta}{1+\Theta} \left(
\frac{1-\Theta}{1+\Theta} \right)^n \;.
\end{equation}
The spatial representation of the
eigenfunctions is 
\begin{eqnarray}\label{hermite}
\varphi_n(Q+\langle q\rangle,t) &=&
(\pi \sigma\Theta)^{-1/4} (2^n n!)^{-1/2}
H_n(Q/\sqrt{\sigma\Theta})\\ \nonumber
&\times& \exp\left(
\frac{i}{\hbar}\frac{m^3\dot{\sigma}(d\langle q \rangle/dt)^2}{
8 \langle [\Delta p]^2\rangle } \right)  \\ \nonumber
&\times&
\exp \left( - \frac{Q^2}{2\sigma\Theta} + \frac{im Q}{\hbar}
\left[  \frac{\dot{\sigma}Q}{4\sigma} + \frac{d\langle q\rangle}{dt}
 \right]\right) \;,
\end{eqnarray}
where $H_n$ are the usual Hermite polynomials.
Transforming also $\widetilde{H}$ back into the original basis,
we obtain
\begin{equation}\label{ham}
H(t) = \frac{(\Delta p - m \dot{\sigma} \Delta q/2\sigma)^2}{2m}
+ \frac{m\widetilde{\Omega}^2[\Delta q]^2}{2} \;.
\end{equation}
In the end, the density operator can be written in the
coordinate-independent form
\begin{equation}\label{densop}
\rho(t) = \widetilde{Z}^{-1} \exp\left(-\widetilde{\beta}(t)
H(t) \right)\;.
\end{equation}
Using Eq.~(\ref{eigen}), one readily checks that
\[
{\rm Tr}\,[\rho(t)] = \sum_{n=0}^\infty \lambda_n(t) = 1 \;,
\]
and similarly one recovers Eq.~(\ref{den2}), since
\[ 
{\rm Tr} \,[\rho^2(t)] = \sum_n \lambda_n^2(t) = \Theta(t) \;.
\]
The linear entropy is then given by
\begin{equation}\label{linent}
S_{\rm lin}(t) \equiv 1 - {\rm Tr}\, [\rho^2] = 1 - \Theta(t)  \;,
\end{equation}
and the Shannon entropy is
\begin{eqnarray}\label{shannon}
S (t)& \equiv& - {\rm Tr}\,[\rho \ln \rho]\\ \nonumber
&=& - \sum_n \lambda_n \ln \lambda_n \\ \nonumber
&=& \frac12 \ln\left(\frac{1-\Theta^2}{4\Theta^2}\right)+
\frac{1}{2\Theta} \ln\left(\frac{1+\Theta}{1-\Theta}\right) \;.
\end{eqnarray}

\section{Microscopic system-plus-bath model} \label{theory}

Let us now consider a wave packet moving in a harmonic
potential surface under the influence of a bath composed of harmonic
oscillators.  The harmonic oscillator modes need not correspond to
physical modes but could represent effective modes chosen to mimic the actual
environment in an optimal way.  Such a system-plus-bath model
allows us to derive exact expressions for the independent 
expectation values $\langle q(t)\rangle$, $\sigma(t)$, and
$\Theta(t)$ of Sec.~\ref{gauss}, and thereby to obtain the exact
quantum dynamics of the damped wave packet for a given
spectral density of the bath modes. 
In this section, the simpler case of a factorized preparation is treated.
The correlated preparation is then discussed in Sec.~\ref{corrp}.

We study a system-plus-bath model \cite{weiss},
$H=H_S+H_B+H_I$, where the ``system'' part describing the 
undamped coherent nuclear motion reads
\begin{equation} 
H_S = \frac{p^2}{2m} + \frac{m\omega_0^2}{2} (q-q_0)^2 \;.
\end{equation}
The ``bath'' is composed of harmonic oscillators coupled linearly  
to the system coordinate,
\begin{equation} \label{bathh}
H_B + H_I = \sum_j \left( \frac{p_j^2}{2m_j}
+ \frac{m_j\omega_j^2}{2} \left[
x_j - \frac{c_j}{m_j \omega_j^2} q \right]^2 \right) \;.
\end{equation}
The  influence of the bath onto the system is fully
specified by the spectral density, 
\begin{equation}\label{spect}
J(\omega) = \frac{\pi}{2} \sum_j \frac{c_j^2}{m_j \omega_j}
\delta(\omega-\omega_j) \;.
\end{equation}
A frequently used model spectral density 
is given by the ohmic bath with a Drude cutoff \cite{weiss},
\begin{equation}\label{drudes}
J(\omega) = \frac{m\gamma \omega}{1+(\omega/\omega_D)^2} \;.
\end{equation}
Under the factorized preparation, the density
matrix at time $t=0$  factorizes according to
\begin{equation} \label{factorize}
\rho(t=0)=\rho_S(q) \otimes \rho_B(\{x_j\}) \;.
\end{equation}
Here $\rho_S(q)$ describes a pure Gaussian
wave packet of width $\sigma_0$ centered
around $q=0$, corresponding to the wavefunction
\[
\psi(q) = (\pi \sigma_0)^{-1/4} \exp(-q^2/2\sigma_0) \;.
\]
The bath is assumed to be in a thermal distribution, 
with the system coordinate held fixed at $q=0$.
While equilibrium properties of a damped harmonic oscillator have been studied 
exhaustively in the past, the consideration of wave-packet 
initial preparations and their corresponding time evolution
leaves room for our contribution.

Due to the harmonic nature of the total system-plus-bath
complex, the exact time-dependent density matrix $\rho(q,q',t)$ 
can then be directly obtained from Feynman-Vernon theory
\cite{weiss,grabert88}.  Switching to symmetric and 
antisymmetric linear combinations,
\begin{equation}\label{comb}
x=(q+q')/2 \;,\quad y=q-q' \;,
\end{equation}
the propagating function of Ref.\cite{grabert88} immediately
leads to the result
\begin{eqnarray} \label{exac}
\rho(x,y,t)& =& \frac{1}{\sqrt{\pi\sigma}} \exp
\Bigl \{ - (x-\langle q\rangle)^2/\sigma - y^2/4\sigma \Theta^2 \\
\nonumber &+& \frac{i}{\hbar} m y  \left(
\frac{\dot{\sigma}(x-\langle q\rangle)}{2\sigma} + \frac{d\langle
q\rangle}{dt} \right) \Bigr \} \;,
\end{eqnarray}
in accordance with the general form (\ref{rhodef}). Now the three independent
expectation values can be expressed in terms of microscopic parameters,
\begin{eqnarray}\label{qdef}
\langle q(t) \rangle &=& \omega_0^2 q_0 \int_0^t dt' \, G(t') \;,\\
\label{sigma1}
\sigma(t) &=& \sigma_0\dot{G}^2(t) +
\frac{\hbar^2}{m^2\sigma_0} G^2(t) + \frac{2\hbar}{m^2} K_q(t) \;,
\end{eqnarray}
and  $\Theta(t)$ is defined by Eq.~(\ref{theta}) with
\begin{equation} \label{p1}
\langle [\Delta p]^2 \rangle=
\frac{m^2}{2} \left( \sigma_0 \ddot{G}^2(t) + \frac{\hbar^2}{m^2\sigma_0}
\dot{G}^2(t) + \frac{2\hbar}{m^2} K_p(t) \right) \;.
\end{equation}
The definition of the functions $G(t)$, $K_q(t)$, and $K_p(t)$ 
for an arbitrary spectral density $J(\omega)$ can be found in the
appendix.  From these expressions, one verifies that the correct
equilibrium values $\sigma_\beta$ and $\langle [\Delta p]^2 \rangle_\beta$
\cite{weiss} are approached at long times.

\begin{figure}
\epsfxsize=0.8\columnwidth
\hspace{.4cm}\epsffile{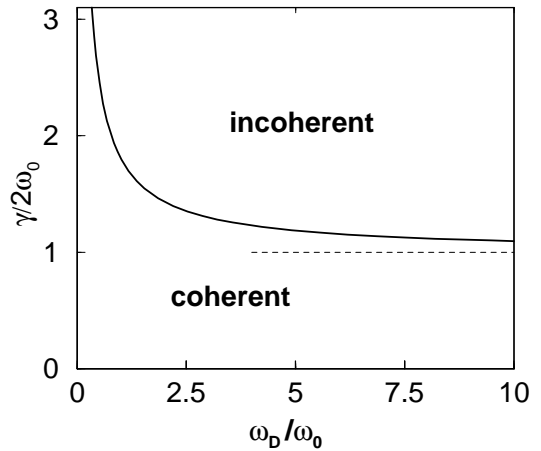}
\caption[]{\label{fig1} Critical damping strength $\gamma_c$ as a function of
$\omega_D$ (solid curve). The limiting value $\gamma_c/2\omega_0=1$ for
$\omega_D\gg \omega_0$ is indicated by the dashed line.}
\end{figure}

Using the ohmic spectral density (\ref{drudes}), we now discuss the question
of coherence of the damped wave packet.  Above a critical value $\gamma_c$ of
the damping strength $\gamma$, where $\gamma_c$ follows from Eq.~(\ref{det}),
oscillations in $\langle q(t) \rangle$ disappear and only incoherent
relaxation can take place, see Figure \ref{fig1}.  For $\omega_D \gg
\omega_0$, the critical damping strength is given by $\gamma_c=2\omega_0$
\cite{weiss}.  While this limit is of most interest in solid-state
applications, the regime $\omega_D\approx \omega_0$ as well as $\omega_D\ll
\omega_0$ has many applications in chemical systems.  Interestingly, the
value of $\gamma_c$ increases when $\omega_D/\omega_0$ becomes small.  In
fact, for $\omega_D \to 0$, the dynamics is always fully coherent,
$\gamma_c=\infty$.  In that limit, the bath is too slow to cause relaxational
behavior.  It is n\-oteworthy that the precise value of $\gamma_c$ depends on
the quantity considered in defining coherence.  Taking the disappearance of
the inelastic peaks in the spectral function as the relevant criterion leads
to a critical value that is smaller by a factor $1/\sqrt{2}$ \cite{pre}.
Since our coherence criterion is based \linebreak[4]
\noindent
on the oscillatory behavior of
$\langle q(t)\rangle$, where the latter equals the corresponding expression for a point-like particle, the
coherent-to-incoherent transition occurs at the same damping strength
$\gamma_c$ for a wave packet and a point-like particle. In particular,
$\gamma_c$ takes a temperature-independent value.

\begin{figure}
\epsfxsize=0.8\columnwidth
\hspace{.7cm}\epsffile{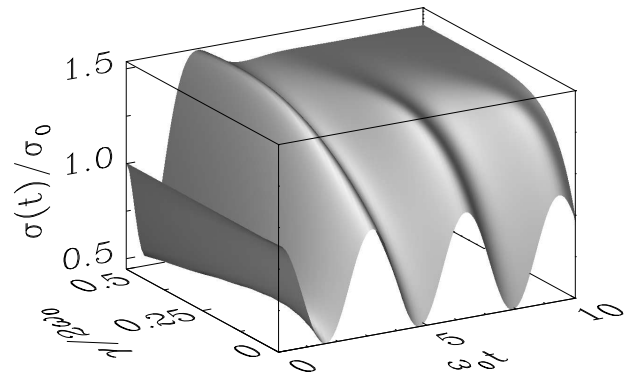}
\vspace{3mm}
\caption[]{\label{fig2} 
Variance  $\sigma(t)$ as a function of $\gamma$ for $\hbar \beta \omega_0=1$,
$\sigma_0=1.5\,\hbar/m\omega_0$, and  
$\omega_D/\omega_0=5$. }
\end{figure}

\begin{figure}
\epsfxsize=0.8\columnwidth
\hspace{.7cm}\epsffile{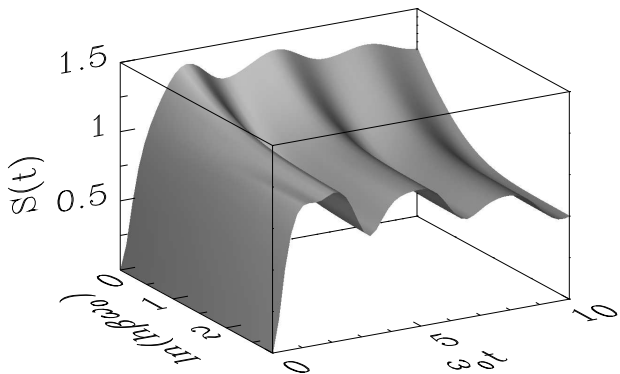}
\vspace{3mm}
\caption[]{\label{fig3} 
Shannon entropy $S(t)$ as a function of $\hbar\beta\omega_0$
for $\gamma/2\omega_0=0.1$,
$\omega_D/\omega_0=15$, and $\sigma_0=1$. }
\end{figure}

Next we briefly discuss the time dependence of the variance $\sigma(t)$,
see Figure \ref{fig2}, and of the Shannon entropy
$S(t)$, see Figure \ref{fig3}. The initial width $\sigma_0$ of the 
wave packet \linebreak[4] 
\noindent
mainly influences the dynamics during the initial stage
of the relaxation. Expanding for small times $\delta t$, the variance reads
\begin{equation} \label{sigma_entwicklung}
\sigma(\delta t) \simeq \sigma_0 + \left[ 
\frac{\hbar^2} {m^2\sigma_0} - \sigma_0(\omega^2 + \gamma\omega_D)
\right] \delta t^2 \;.
\end{equation}
Therefore the variance initially increases (decreases) 
for $\sigma_0< \tilde{\sigma}$  ($\sigma_0>\tilde{\sigma}$), where  
$\tilde{\sigma}=\hbar/m\sqrt{\omega_0^2 + \gamma\omega_D}$.  
For $\gamma<\gamma_c$, oscillations in both $\sigma(t)$
and $S(t)$ are found,  similar to the behavior of
$\langle q(t) \rangle$. These oscillations
again persist at high temperatures, albeit with smaller
ampli-\linebreak[4]\pagebreak
\vspace{4mm}
\noindent
tude.
The initial entropy increase observed in Fig.~\ref{fig3} 
becomes very pronounced if $\sigma_0$ strongly deviates 
from the natural width $\sigma_\beta$ of the damped oscillator. 
Furthermore, initial transient oscillations
then persist for a longer time.  They are particularly pronounced 
for low temperatures and small $\sigma_0$,  with
a transient entropy  large compared to the equilibrium value $S(t\to \infty)$.

\section{Pump-probe spectroscopy of the Reaction Center}
\label{vossec}

\begin{figure}
\epsfxsize=0.8\columnwidth
\hspace{.4cm}\epsffile{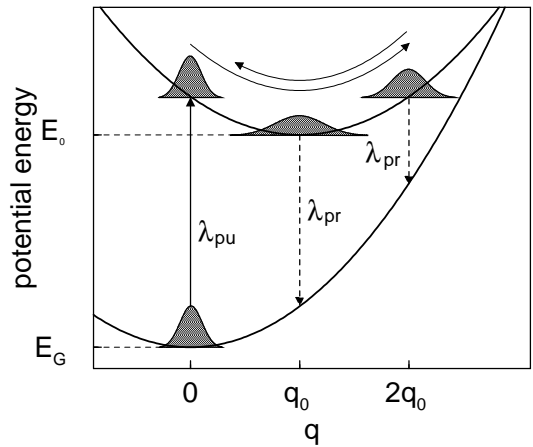}
\caption[]{\label{fig4} 
Pump-probe setup involving two harmonic surfaces. The
dark (excited) state surface is centered at $q=0 \; (q=q_0)$.
}
\end{figure}

Next we apply the results presented before in a specific context. The system
under study is the photosynthetic reaction center in purple bacteria. In recent
pump-probe experiments on modified and wild-type reaction centers, Vos {\em et
al.} \cite{vos93,vos94,vos98} have observed oscillations in the time-resolved
emission signal, which were interpreted to reflect coherent nuclear motion in
the excited electronic state (``vibrational coherence'').  This observation
immediately received much attention, as coherent dynamics was not expected to
exist in such a condensed-phase system.  Clearly, a nuclear coordinate within a
macromolecule like the reaction center could be drastically influenced by
dissipation, which suggests a treatment similar to the one discussed above.
The situation that Vos {\em et al.} constructed from their data is depicted in
Fig.~\ref{fig4}. The excited state surface was found to be parabolic with a
curvature of $\omega_0 = 75 \, {\rm cm}^{-1}$. It was populated with a 870 nm
pump pulse, say, at time $t=0$, and probed with pulses around 921.5 nm,
corresponding to the minimum of the excited state surface.  In this section, we
expand on the above analysis in order to describe the emission signal. Thereby
effects of the spectral density characteristics and of the initial correlations
can be captured, where especially the latter are missed by any simpler
formalism. In order to clearly show these initial correlation effects, we
shall crudely simplify the modelling of the pump (and to a lesser extent of the
probe) pulse. In particular, we \linebreak[4]
\noindent
make the (strictly speaking unphysical) assumption that the pump pulse
transfers the complete nuclear wave packet up to the excited state
surface. Therefore we only have to treat the dissipative excited state dynamics
up to the probe pulse. Of course, thereby potentially important effects like
the impulsive resonance Raman contribution \cite{domcke,mukamel,wolf} are
missed. However, in principle our theory can straightforwardly be extended to
provide a more realistic modelling of the pump and probe processes.

\subsection{Model and parameters}

The Hamiltonian $H(t) = H_0 + V(t)$ governing the emission process first
consists of an unperturbed Hamiltonian
\begin{equation} \label{H_0}
H_0 = |G\rangle H_G \langle G|  + |E\rangle ( H_E  + H_I + H_B) \langle E| \; .
\end{equation}
The orthonormal states $|G\rangle$ and $|E\rangle$ denote the electronic
degrees of freedom, with $H_G$ ($H_E$) being the Hamiltonian in the ground
(excited) state,
\begin{eqnarray*}
H_G & = & \frac{p^2}{2m} + \frac{m\omega_G^2}{2} q^2 \;, \\
H_E & =& \frac{p^2}{2m} + \frac{m\omega_0^2}{2} (q - q_0)^2 + 
\hbar\omega_\Delta \;,
\end{eqnarray*}
where $\hbar\omega_\Delta=E_0-E_G$ and $q_0$ is the separation of the potential
minima, see Fig.~\ref{fig4}.  The dissipation acting on the wave packet in the
excited state is included via $H_I+H_B$, see Eq.~(\ref{bathh}). Since we only
consider the dynamics on the excited state surface {\sl between} the pump and
the probe pulse, it is not necessary to account for dissipation in the ground
state at $t>0$. The effect of the probe pulse is described by $V(t)$. Under the
dipole and the rotating wave approximation \cite{mukamel},
\begin{equation} \label{V(t)}
V(t) = \epsilon(t)  |G\rangle\langle E|  + \; {\rm H.c.} \;,
\end{equation}
where $\epsilon(t)$ represents the temporal envelope of the electric field.
The probe pulse was taken in the form
\begin{equation} \label{pulse}
\epsilon(t') = \theta[t'-(t-\delta)]\,
 \theta[t+\delta-t'] \,e^{i\omega_{pr} t'} \;,
\end{equation}
where $\theta$ is the Heaviside function and the
probe pulse is centered at time $t$. 
Since the 30 fs probe pulses used
in Ref.~\cite{vos93} did not maintain their full intensity over the whole pulse
duration, we have chosen a smaller duration of $2\delta=20$ fs. 

The model parameters were taken as follows. 
 The frequency of the excited
[ground] state surface is $\omega_0=75~{\rm cm}^{-1}$
 [$\omega_G=130~{\rm cm}^{-1}$]. Furthermore, $q_0$ and $\hbar\omega_\Delta$
are calculated from the wavelength of the pump pulse, $\lambda_{pu}=870$~nm,
and of the probe pulse acting at $q=q_0$, 
$\lambda_{pr}=921.5$~nm.  The parabolic geometry of Fig.~\ref{fig4} then
yields $q_0=2.07 \sqrt{\hbar / m \omega_0}$ 
and $\omega_\Delta=11\,327$ cm$^{-1}$.
At this point, little is known about microscopic details of the dissipation
acting on the reaction coordinate $q$. 
In principle, one should first compute
the appropriate spectral density for the system under consideration
by means of MD simulations \cite{david}.  
In the absence of such information, we make the assumption
of an ohmic bath with a Drude cutoff, see Eq.~(\ref{drudes}).
This spectral density was shown to be in agreement with the
overall structure of the spectral density coupling to the
primary electron transfer step in the reaction center \cite{david}.
To account for the lack of knowledge concerning the spectral density, we
have studied two different spectral parameter sets. The first one, which is
referred to as SP~I, is $\gamma/2\omega_0=0.1$ and $\omega_D/\omega_0=100$.
The second one, referred to as SP~II, is $\gamma/2 \omega_0=0.75$ and
$\omega_D/\omega_0=0.5$.  Both sets are within the coherent regime,
$\gamma<\gamma_c(\omega_D)$, and are chosen such that the oscillations in
$\langle q(t)\rangle$ decay on the same time scale as those of the $T=10$ K
emission signal reported in Ref.~\cite{vos93}.

\subsection{Initial preparation}
\label{corrp}

A conceptually more severe point concerns the proper description of the initial
state ($t=0$).  Again, two very different
initial preparations are conceivable. The first one is to assume a wave packet
in the usual sense, where the oscillator is initially in a pure state without
correlations with the bath.  This is the ``factorized preparation'' elaborated
in Sec.~\ref{theory} and (at least implicitely) employed in most previous
treatments.  On the other hand, the nuclear coordinate already experiences the
environment while the system is in the ground state, and therefore the initial
density matrix does not factorize.  A more realistic preparation is to take the
$|G\rangle$ oscillator at equilibrium with the same bath as in the excited
state, whence there will be system-bath correlations at $t=0$ (``correlated
preparation'').  We mention in passing that the correlated preparation is
related to the initial bath preparation discussed in Ref.~\cite{lucke} in the
context of electron transfer reactions. The special case $\omega_G=\omega_0$
with $\omega_D \gg \omega_0$ has also been treated in 
Ref.~\cite{mukamel} and references therein.

Due to its very short duration, as a result of the pump pulse at $t=0$,
the system is assumed to suddenly change from the ground state to the excited
state surface. This
amounts to both a vertical shift and a change in curvature $\omega_{G}
\rightarrow \omega_0$, see Fig.~\ref{fig4}. Technically speaking, the
correlated preparation can be most conveniently accounted for by following the
path-integral analysis of Ref.\cite{grabert88}, but keeping different system
potentials acting on the imaginary-time and real-time paths.  The resulting
reduced density matrix is then of the form (\ref{exac}) again. Due to the
Ehrenfest theorem, $\langle q(t)\rangle$ and $\langle p(t)\rangle$ coincide
with the results of the factorized preparation.  The variances $\sigma(t)$ and
$\langle [\Delta p]^{2} (t) \rangle$ follow in closed form and are given in the
appendix.  For the corresponding results under the factorized preparation, see
Eqs.~(\ref{sigma1}) and (\ref{p1}).

For both preparations, the initial width $\sigma(t=0)$ was chosen as the
thermal width $\sigma_\beta^G$ in the ground state oscillator.  Importantly,
despite of having the same initial value, the time-dependence of the variance
is strikingly different depending on the initial condition.  This becomes
particularly evident for $\omega_{G} = \omega_0$, where for the correlated
preparation, $\sigma(t)$ and $\langle [\Delta p]^{2} (t) \rangle$ stay constant
in time, whereas the factorized preparation {\sl always} leads to
time-dependent variances.  This can be understood by noting that $\langle
[\Delta p]^2(t=0) \rangle$ for the factorized preparation is determined by the
minimum uncertainty condition $\Theta(t=0)=1$, see Eq.~(\ref{theta}), while it
is given by $\langle [\Delta p]^2 \rangle_\beta^G$, see Eq.~(\ref{pp2}), in the
case of a correlated preparation.  Since the deviation in $\langle [\Delta
p]^2(t=0) \rangle$ for the two initial preparations becomes larger with
increasing temperatures, one expects that the choice of the correct initial
preparation is more important at high temperatures.  This is indeed confirmed
by the results for the stimulated emission signal reported below.

\subsection{Calculating the emission signal} \label{emsig}

Next we calculate
the time-dependent total stimulated emission signal.
After the pump pulse at $t=0$, the system is assumed
to be in the excited state surface according to a properly
chosen initial preparation.
The probe pulse is then assumed to be much faster than typical solvent time
scales such that the environmental
 influence can be neglected during the emission
process itself. The time-dependent emission signal can thus be
expressed in terms of
the reduced density matrix directly before and after the application of the
probe pulse.  
For a probe pulse centered at time $t$ with duration $2\delta$, the energy
$E(t)$ emitted during the transition is
\begin{eqnarray} \label{S}
E(t) &=& \langle H_0\rangle_{\rho(t + \delta)} - \langle H_0
\rangle_{\rho(t - \delta)} \\
 \nonumber
 &=& {\rm Tr}\{H_0 [ \rho(t + \delta) - \rho(t - \delta) ]\} \;,
\end{eqnarray}
with the reduced density matrix $\rho(t)$.  Herein the influence of the
bath during the emission process has been neglected.
Adopting a matrix representation for $\rho(t)$ with respect to the electronic
states $|G\rangle$ and $|E\rangle$, we notice that $\rho(t' < t - \delta) =
|E\rangle \rho^E(t') \langle E|$, since for $t' < t - \delta$, the wave
packet is located on the excited state surface. For $t' > t - \delta$,
however, $V(t')$ causes a population of other matrix elements as well.  Since we
are interested in the emission signal, the trace in Eq.~(\ref{S}) allows us to
focus only on the diagonal elements. Using second-order perturbation theory in
$V(t)$, they read
\begin{eqnarray} \label{rhoconts}
\rho^G(t + \delta) &=& U_{1,GE}(t + \delta, t - \delta) \,
 \rho^E(t - \delta)\\ 
\nonumber
&\times& U_{1,EG}^{-1}(t + \delta, t - \delta) \;, \\
\nonumber
\rho^E(t + \delta) &=& U_0(t + \delta, t - \delta) \, \rho^E(t - \delta)
   \, U_0^{-1}(t + \delta, t - \delta)  \\ 
\nonumber
&+& \Big[ U_2(t + \delta, t - \delta) \, \rho^E(t - \delta)
   \, U_0^{-1}(t + \delta, t - \delta) \\ 
\nonumber
&& \quad +\; {\rm H.c.} \Big]          \;.
\end{eqnarray}
Here $U_k(t,t')$ denotes the appropriate matrix element of the $k$th term of
the Dyson expansion for the time evolution operator under $H(t)$,
\begin{eqnarray*} 
U_0(t, t') &=& e^{ -\frac{i}{\hbar} H_E (t - t') } \,, \\
U_{1,EG}(t, t') &=& - \frac{i}{\hbar} \int_{t'}^{t} dt_1
\epsilon^\ast(t_1) e^{ -\frac{i}{\hbar} H_E (t - t_1) } 
e^{ -\frac{i}{\hbar} H_G (t_1 - t') } \, ,  \\
U_2(t, t') &=& - \frac{1}{\hbar^2}
\int_{t'}^t\, dt_1 \int_{t'}^{t_1}\,dt_2\,\epsilon(t_1) \epsilon^\ast(t_2) 
e^{ -\frac{i}{\hbar} H_E (t-t_2) }  \\
&\times& e^{ -\frac{i}{\hbar} H_G(t_2-t_1) } e^{ -\frac{i}{\hbar}
 H_E(t_1-t') }  \;,
\end{eqnarray*}
with $U_{1, GE}(t, t') = - U_{1, EG}^\dag(t, t')$.  After some algebra, we
obtain the time-resolved total emission signal in the form
\begin{eqnarray} \label{S3}
E(t) &=& \frac{1}{\hbar}\,\sum_{n,r,s=0}^\infty \langle n | r \rangle
\langle s|n\rangle \Bigg( 2[\omega_0 (s +1/2) + \omega_\Delta]
\\ \nonumber &\times&
{\rm Re} \Big\{ \rho_{rs}^E(t - \delta) e^{i\omega_0(r-s)(t-\delta)}  
\\ \nonumber &\times&
 \int_{t-\delta}^{t+\delta} \,dt'\, \epsilon(t')
\int_{t-\delta}^{t'}\, dt''\, \epsilon^\ast(t'')
\\ &\times& \nonumber \exp \{i\,[ \omega_G ( n + 1/2 ) 
- \omega_0 ( r + 1/2 ) -\omega_\Delta ]t'\} \\ \nonumber
 &\times& \exp\{i\, [ \omega_0 ( s + 1/2) + 
\omega_\Delta - \omega_G ( n + 1/2 ) ]t'' \} \Big\} 
  \\ \nonumber &-& \omega_G\, (n + 1/2)\rho_{rs}^E(t - \delta)
e^{i\omega_0(r-s)(t-\delta)}
\\ \nonumber &\times&
\int_{t-\delta}^{t+\delta}\,dt'\,
      \epsilon(t') \int_{t-\delta}^{t+\delta}\,dt'' \, \epsilon^\ast(t'') 
\\ &\times& \nonumber \exp\{ i\,[ \omega_G  ( n + 1/2 ) - \omega_0
 ( r + 1/2 ) -\omega_\Delta ] t'\} \\ &\times&
\exp \{ i\, [ \omega_0 ( s + 1/2 ) 
+\omega_\Delta - \omega_G ( n + 1/2 ) ]t''\} \Bigg)
 \;, \nonumber
\end{eqnarray}
where $|n\rangle$ and $|r,s\rangle$ denote the vibronic eigenstates of $H_G$
and $H_E$, respectively.

In principle, the above analysis can straightforwardly be extended
in order to incorporate the pump pulse.
The resulting initial reduced density matrix
is then composed of four different contributions, namely those in
Eq.~(\ref{rhoconts}) and the two nondiagonal terms. The subsequent
time evolution with both electronic surfaces coupled to the bath could then be
treated in a similar way as presented in Sec.~\ref{theory}.

\subsection{Results for the reaction center}

\begin{figure}
\epsfxsize=0.8\columnwidth
\hspace{.4cm}\epsffile{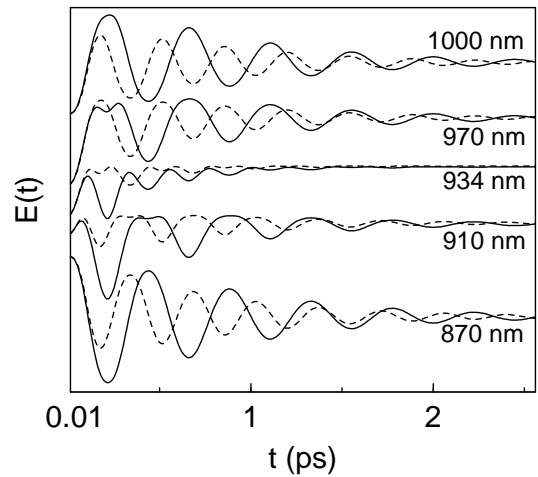}
\caption[]{\label{fig5}
Emission signal $E(t)$ [in arbitrary units] for different probe wavelengths
$\lambda_{pr}$ at $T=10$ K for the factorized preparation. The
solid (dashed) curve is for SP~I (SP~II). For clarity,
curves for subsequent values of $\lambda_{pr}$ have been shifted
vertically.
}
\end{figure}

Figure \ref{fig5} shows the time-resolved stimulated emission signal for
different probe wavelengths $\lambda_{pr}$ at $T=10$~K. At such a low
temperature, the difference between the factorized and correlated preparation
is very small and can hardly be resolved in Fig.~\ref{fig5}. The qualitative
features of the emission signal can be understood within the wave-packet
picture by relating $\lambda_{pr}$ to a particular value of the nuclear
coordinate $q$, as is seen by plotting the discrete Fourier-transformed
emission spectrum $A(\lambda_{pr})$ at the frequency $\tilde{\omega}$
corresponding to the ground oscillation, see Fig.~\ref{fig6}. This frequency,
determined from the imaginary part of the roots of Eq. (\ref{cubic}), is 97.2
${\rm cm}^{-1}$ for SP~II but deviates less than 1\% from $\omega_0$ for
SP~I. The maxima in $A(\lambda_{pr})$ then correspond to the left ($q=0$) and
right ($q=2q_0$) turning points of the undamped nuclear wave-packet, while the
minimum is related to the bottom of the potential surface $(q=q_0)$ in
Fig.~\ref{fig4}. Due to the finite pulse duration and the different
Franck-Condon overlap factors for $q>q_0$ and $q<q_0$, the corresponding value
of $\lambda_{pr}$ differs from 921.5 nm, particularly at high temperatures.
Since the turning points are passed once per period but the bottom is visited
twice, the cor-\linebreak[4]
\noindent
responding emission signals should be oscillatory with frequency
$\tilde{\omega}$ and $2\tilde{\omega}$, respectively \cite{vos93}. This
behavior is indeed found in Fig.~\ref{fig5}. Focusing on SP~I, the emission
signal at $\lambda_{pr}=1000$~nm, corresponding to the right turning point,
exhibits a phase shift of $\pi$ and a smaller amplitude compared to
$\lambda_{pr}=870$~nm. This can be explained by noting that the right turning
point is reached half a period later than the left one, whence the most
significant initial contribution is damped more strongly. Apart \linebreak[4]
\begin{figure}
\epsfxsize=0.8\columnwidth
\hspace{.2cm}\epsffile{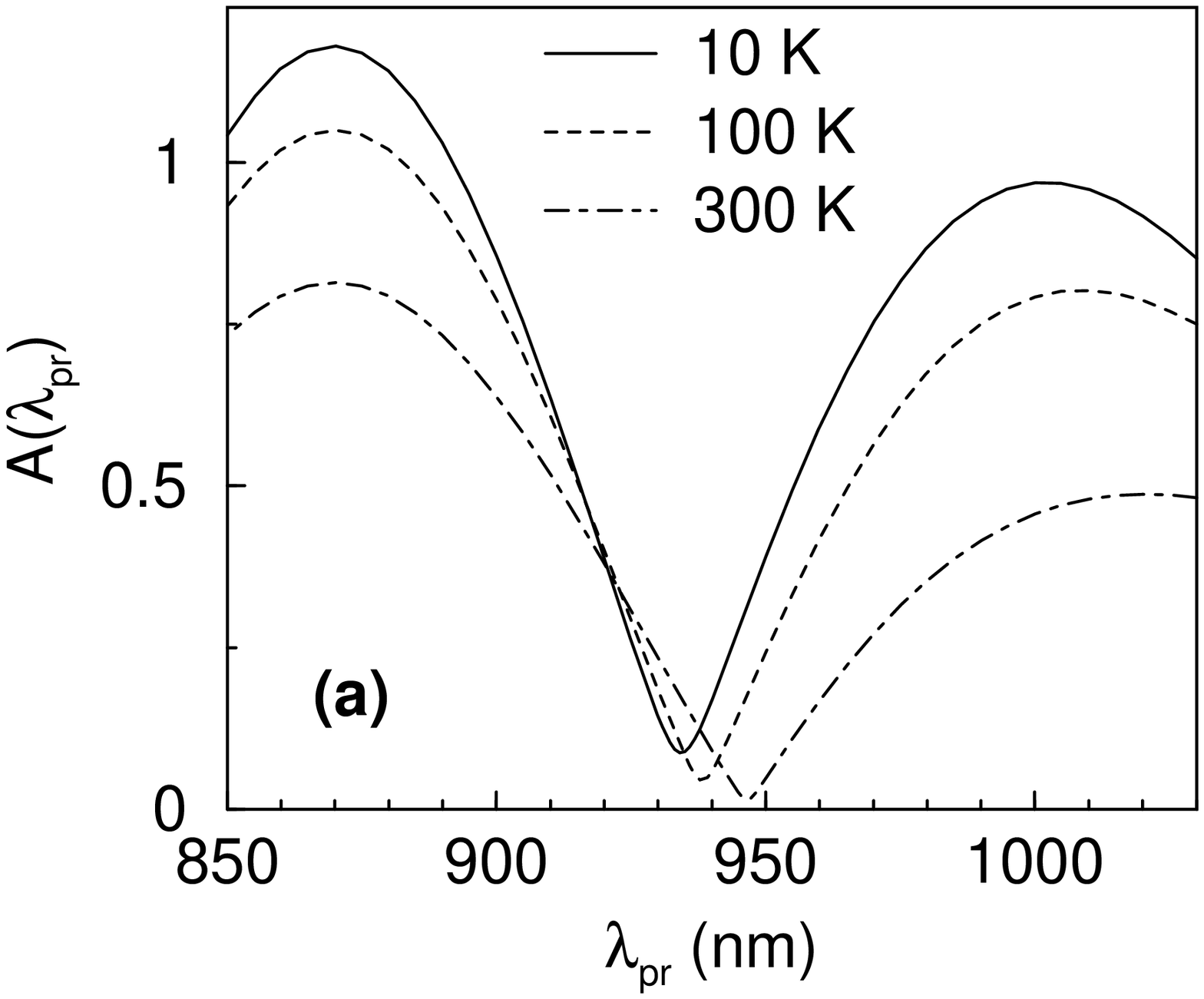}
\end{figure}
\vspace{-.5cm}
\begin{figure}
\epsfxsize=0.8\columnwidth
\hspace{.2cm}\epsffile{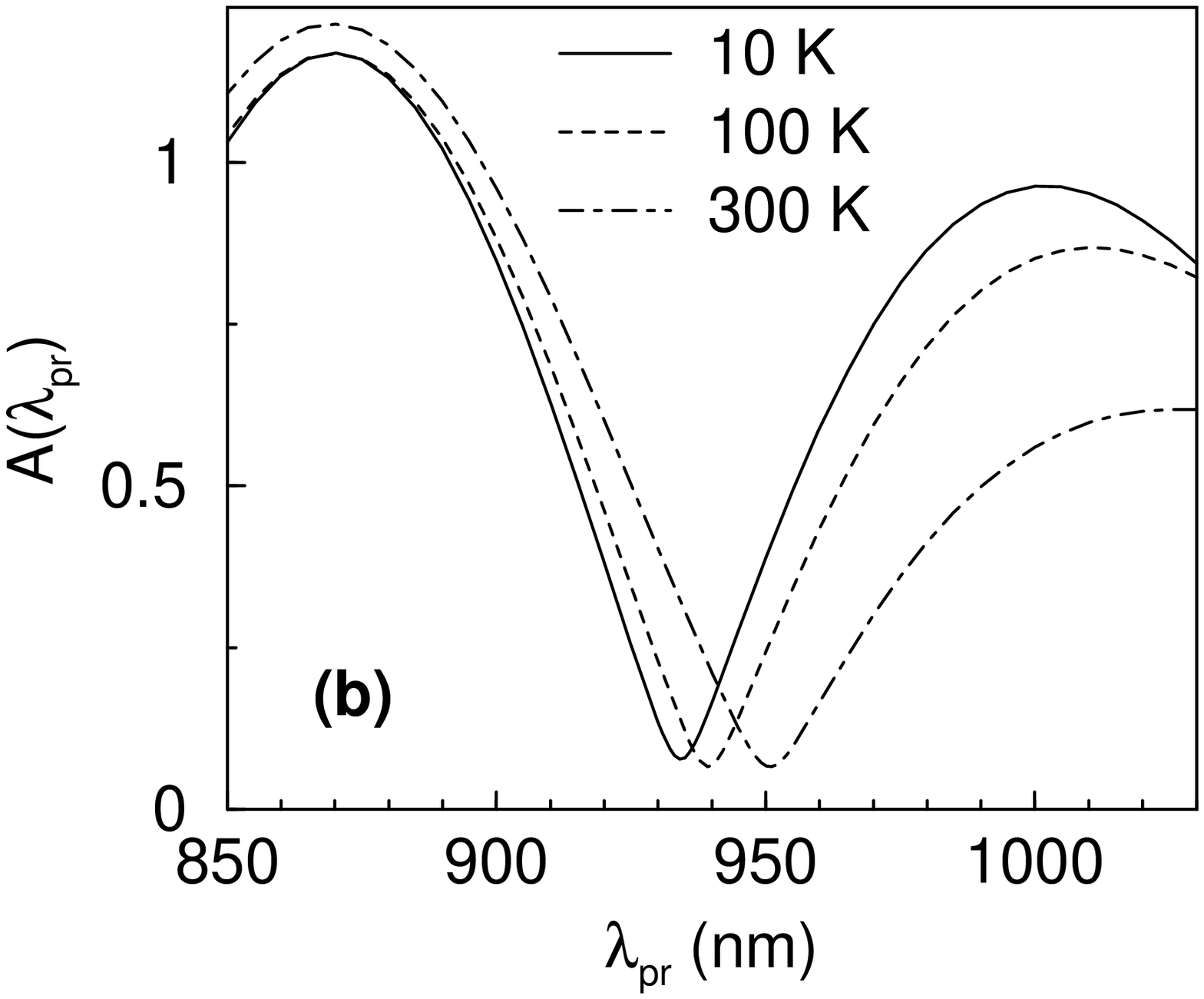}\\
\end{figure}
\vspace{-.9cm}
\begin{figure}
\epsfxsize=0.8\columnwidth
\hspace{.2cm}\epsffile{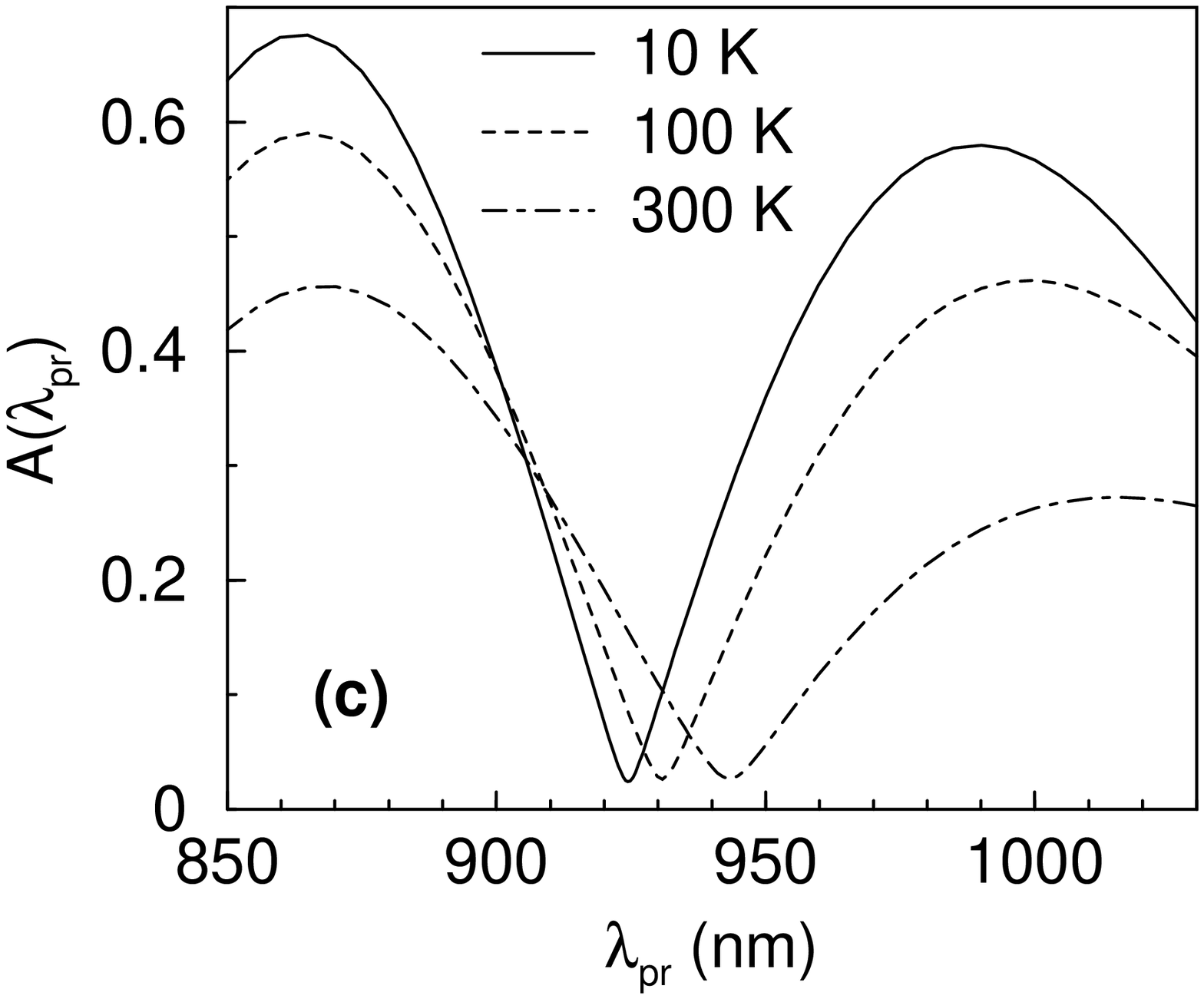}
\end{figure}
\begin{figure}
\epsfxsize=0.8\columnwidth
\hspace{.2cm}\epsffile{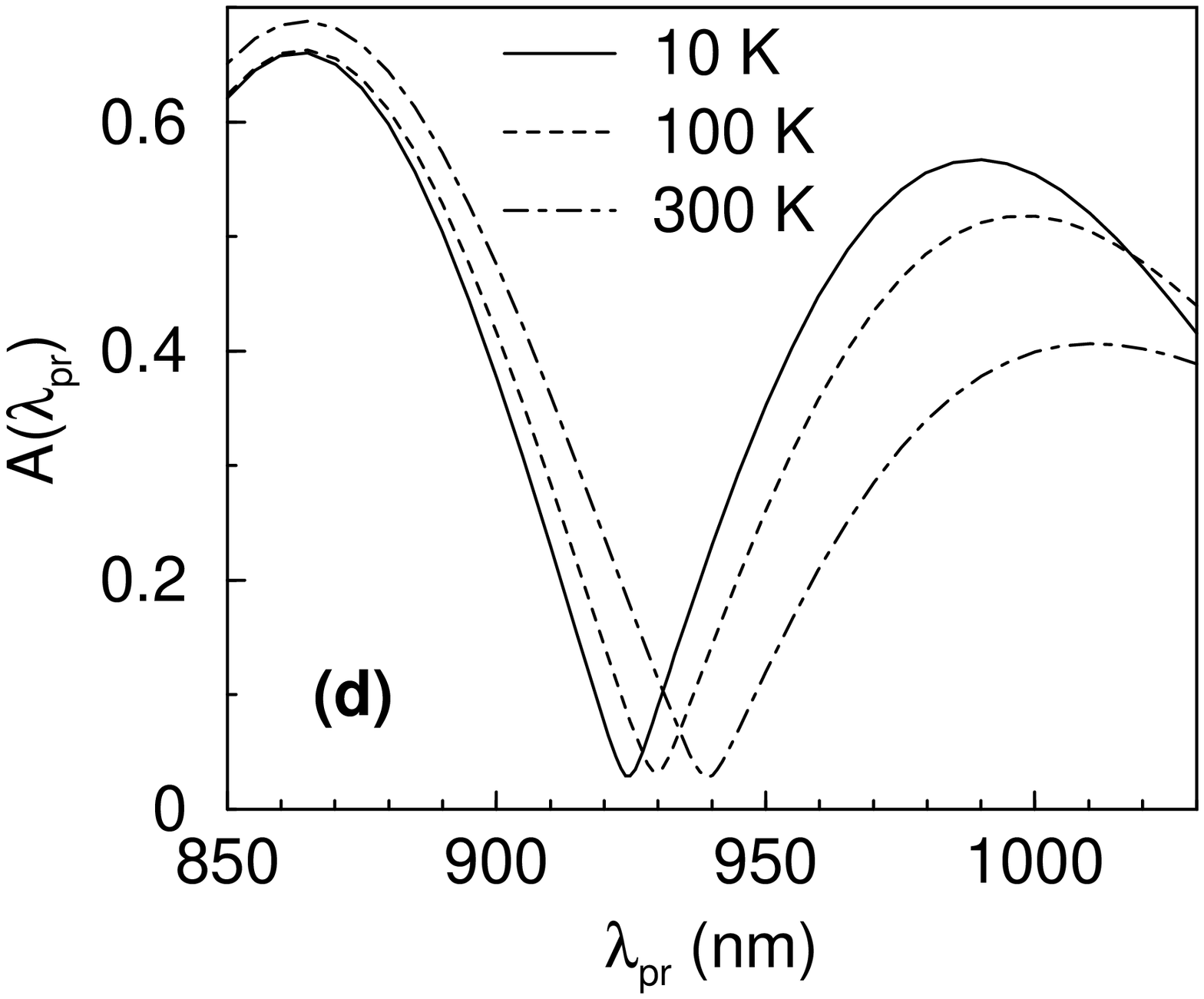}
\caption[]{\label{fig6} 
Emission spectrum $A(\lambda_{pr})$ [in arbitrary units]
at $\tilde{\omega}$ for different temperatures.
In (a) [(b)], we have taken SP~I with a factorized [correlated] preparation. 
In (c) [(d)], we have taken SP~II with a factorized [correlated] preparation.} 
\end{figure}
\noindent
from the decay of the oscillations in $E(t)$, damping of the nuclear motion is
also reflected in a finite amplitude $A(\lambda_{pr}) >0 $ at the minimum. For
wavelengths $\lambda_{pr}$ away from the turning points or the bottom, there
are two different time intervals between subsequent passings of the transition
region. This leads to the splitting of the maxima in the short-time emission
signal observed in Fig.~\ref{fig5}.

With increasing temperature, according to our argumentation above, the
influence of the initial preparation should become more and more important.
This is seen in the temperature dependence of the spectrum and of the emission
signal shown in Figures \ref{fig6} and \ref{fig7}, respectively.  We first
focus on the effects seen in Fig.~\ref{fig6}. The minimum of $A(\lambda_{pr})$
shifts towards higher wavelengths with increasing temperature. This can be
rationalized by noting that for $\omega_0 < \omega_G$, the energy gap between
the $n$th vibronic eigenstates of the excited and ground state decreases with
$n$, and that high-order eigenstates become more important in the spectral
decomposition of $\rho(t)$ at \linebreak[4]
\begin{figure}
\epsfxsize=0.8\columnwidth
\hspace{.4cm}\epsffile{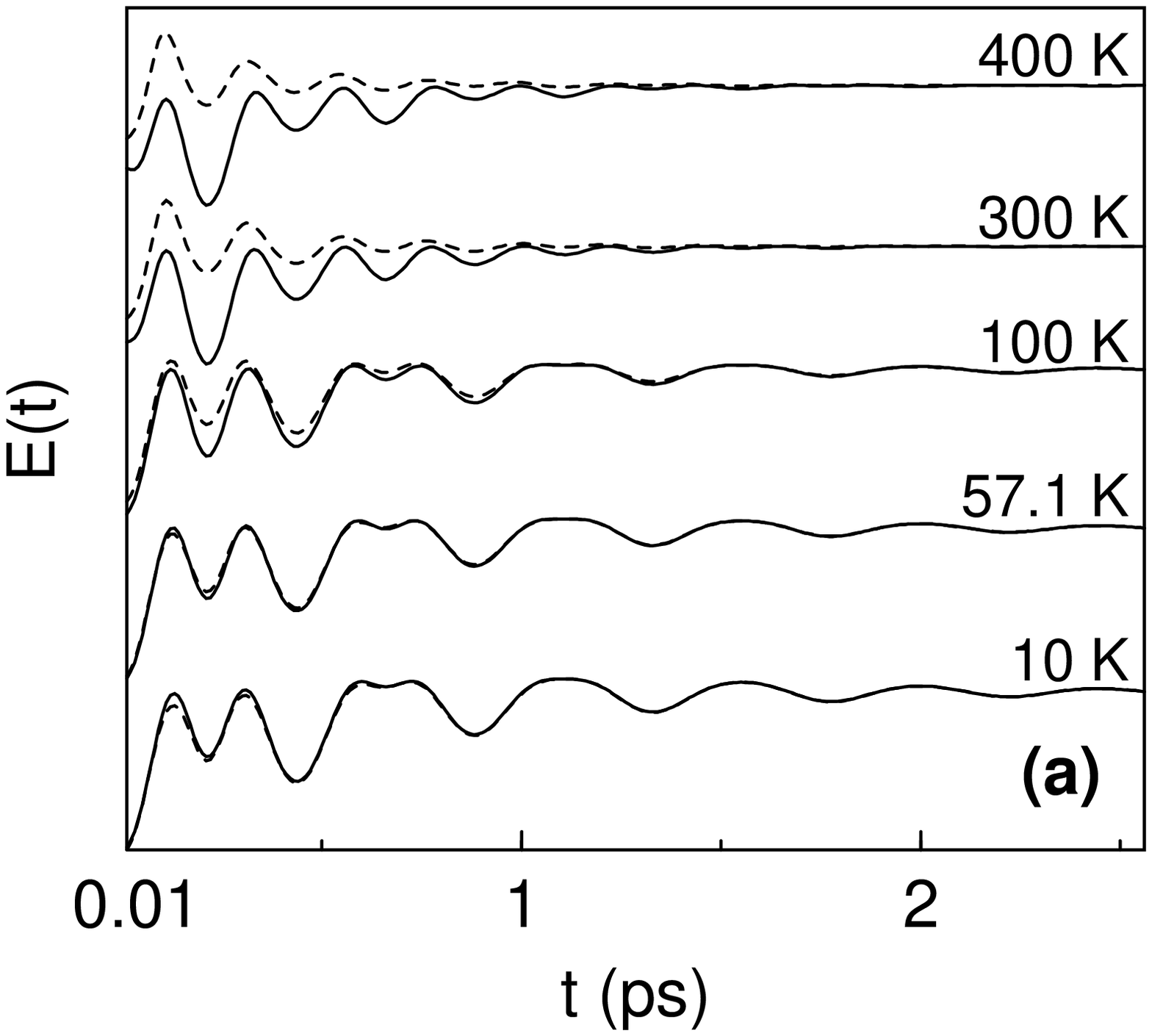}
\end{figure}
\begin{figure}
\epsfxsize=0.8\columnwidth
\hspace{.4cm}\epsffile{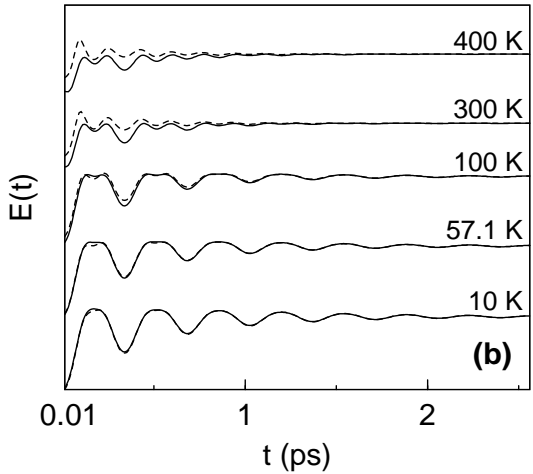}
\caption[]{\label{fig7}
Emission signal $E(t)$ for
$\lambda_{pr}=950$~nm and several temperatures
for (a) SP~I and (b) SP~II. 
The solid (dashed) curve
is for the correlated (factorized) preparation. 
Curves for subsequent temperatures have been shifted
vertically by the same amount.
}
\end{figure}
\noindent
higher temperatures. In fact, additional
calculations for $\omega_G=\omega_0 /2$ [not shown here] yield a similar shift
towards smaller $\lambda_{pr}$. This shift is more pronounced for the
correlated initial preparation, but depends only weakly on the spectral density
of the environmental modes.

\begin{figure}
\epsfxsize=0.84\columnwidth
\hspace{.0cm}\epsffile{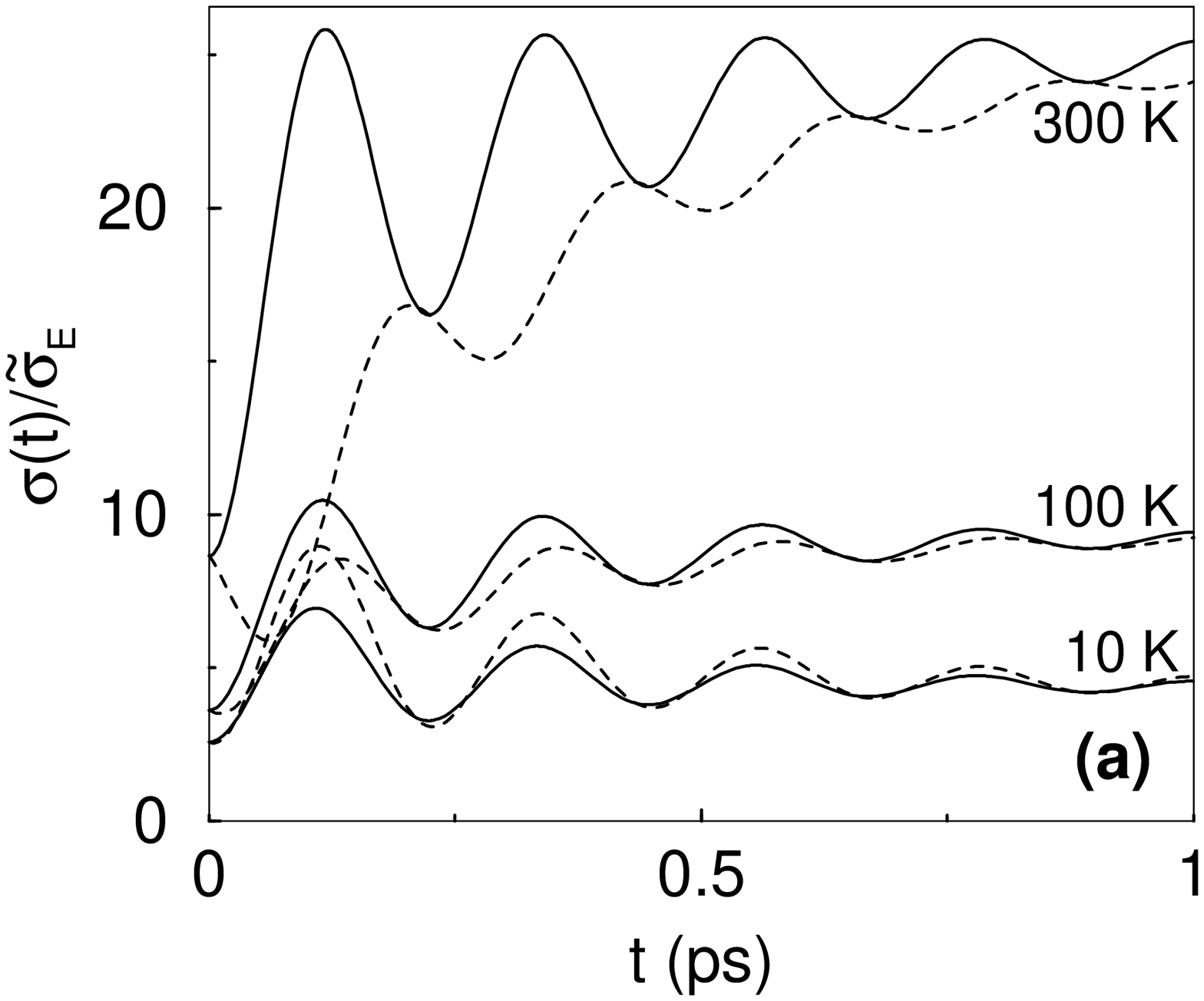}\\
\begin{figure}
\vspace{-1cm}
\end{figure}
\epsfxsize=0.8\columnwidth
\hspace{.3cm}\epsffile{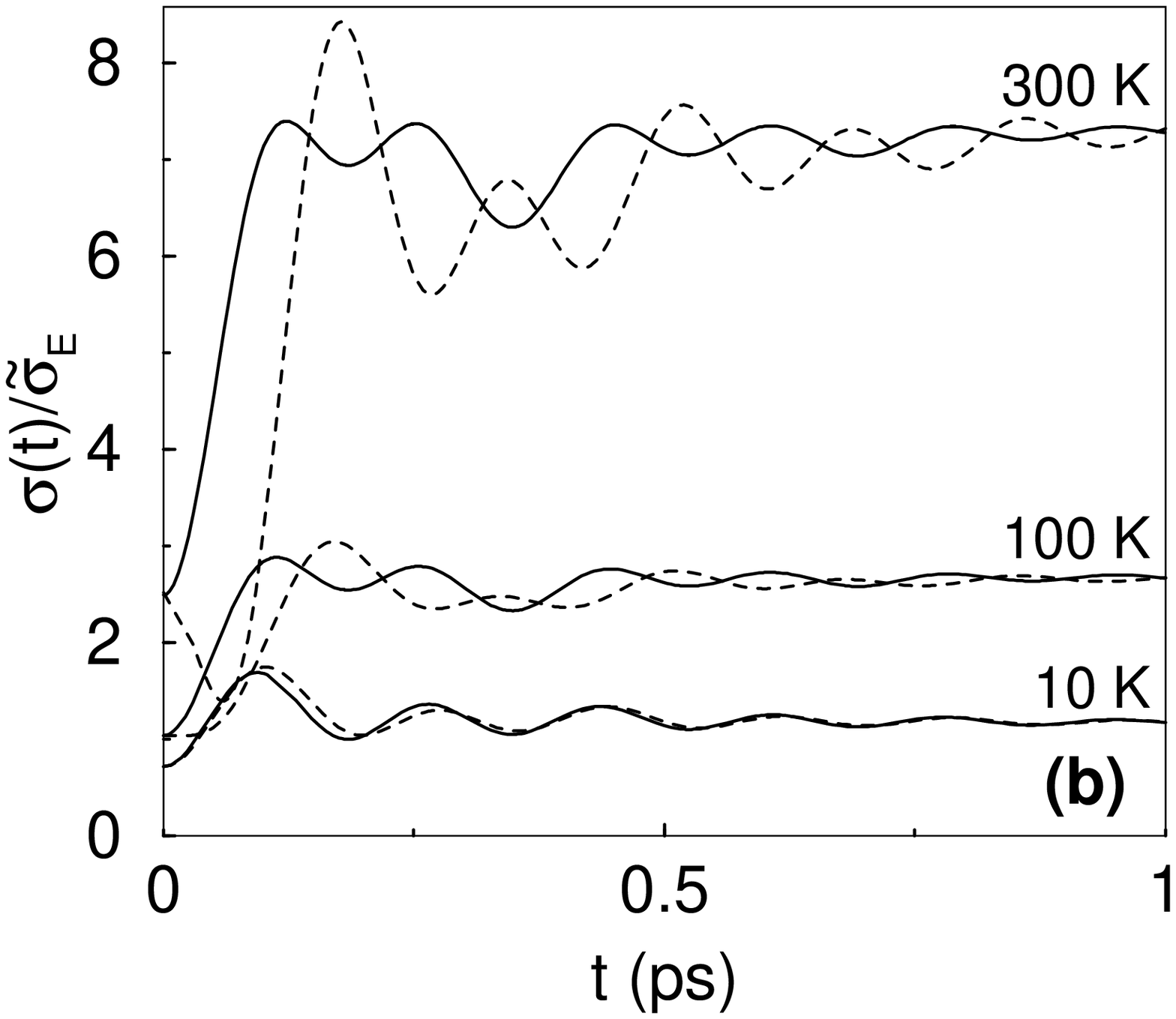}
\caption[]{\label{fig8}
Temperature dependence of $\sigma(t)$ 
[in units of $\tilde{\sigma}_E=\hbar/m\sqrt{\omega_0^2
+\gamma \omega_D}$] for (a) SP~I and (b) SP~II. 
The solid (dashed) curve
is for the correlated (factorized) preparation. 
 }
\end{figure}

Figure 7 shows the temperature dependence of the emission signal at
$\lambda_{pr}=950$~nm, corresponding to $q_0<q<2q_0$, with $q$ approaching the
bottom $q= q_0$ as the temperature is increased.  Furthermore, Figure 8 shows
that $\sigma(t)$ under the factorized preparation experiences a phase shift of
almost $\pi$ at $T>100$~K compared to the correlated preparation.  Such an
unphysical phase shift arises as a relict of the unperturbed evolution of a
harmonic oscillator.  Furthermore, since $\sigma(0)=\sigma_\beta^G$ increases
with temperature, a negative initial slope results under the factorized
preparation, see Eq.~(\ref{sigma_entwicklung}).  Notably, for the correlated
preparation, the maxima in $E(t)$ stay always close to the equilibrium values,
in marked contrast to the factorized preparation but in accordance with the
experimental data of Ref.~\cite{vos93}.  A similar behavior is seen in the
variance shown in Fig.~\ref{fig8}.  For the correlated preparation, the maxima
in $\sigma(t)$ occur every $(k+1/2)$th period, corresponding to the passing of
the bottom $q=q_0$, and they are very close to their equilibrium value
$\sigma_\beta^E$.  Therefore, for the correlated preparation, the independent
expectation values $\langle q(t) \rangle$ and $\sigma(t)$ are always close to
their equilibrium values when passing the bottom $q=q_0$.  On the other hand,
for the factorized preparation, the phase shift in $\sigma(t)$ results in a
bunching of the wave packet when passing the bottom. This causes even
qualitatively different emission signals.  We conclude that at high
temperatures several unphysical effects are introduced by using a factorized
initial preparation, and a wrong description of the emission spectrum may
result.

\section{Conclusions}
\label{conc}

In this work, we have formulated a dissipative wave-packet approach towards a
detailed theoretical description of stimulated emission pump-probe experiments
under condensed-phase conditions. Assuming harmonic surfaces for both the
ground and the excited state, the Gaussian nature of the wave packet describing
the coherent nuclear motion allows for an exact treatment even if strong
damping by environmental modes is present. Modelling the environmental modes
by a set of infinitely many effective harmonic oscillators with a suitably
chosen spectral density, it is then possible to make detailed predictions for
the stimulated emission signal and for the corresponding spectra. While the
spectral density is in principle accessible in terms of MD simulations, we have
studied two model spectral densities in this work. A particular advantage of
our approach is the possibility of treating different initial preparations of
the wavepacket-plus-bath complex directly after the pump pulse ($t=0$). A more
realistic calculation should also explicitely study the pump pulse, which can
in principle be done along the same lines. Under such a formalism, a correct
choice for the initial preparation of the wavepacket-plus-bath complex before
the pump pulse will be important and is expected to lead to similar effects.

The recent experiments by Vos {\em et al.} \cite{vos93,vos94,vos98} on the
bacterial photosynthetic reaction center have been analyzed using this
formalism.  Due to our assumptions about the pump pulse, the possibly important
impulsive resonant Raman contribution was not taken into account here.  While
some of the qualitative features of the coherent nuclear motion have been
discussed before using simpler arguments \cite{vos93}, our approach can allow
for a fully quantum-mechanical comparison of experimental data with
theory. Even in the absence of detailed knowledge about the environmental
spectral density, conclusions of relevance to the interpretation of
experimental results can be extracted from our analysis.  In particular, we
have shown that at high temperatures, the assumption of a factorized initial
state leads to large differences from the theoretical predictions under a more
realistic correlated initial state.

Finally it should be stressed that the approach presented here can be applied
to other pump-probe spectroscopy setups as well.  In particular, if the excited
state surface is weakly coupled to another surface, as happens, e.g., in the
primary electron transfer step in the reaction center, transitions to this
surface are expected to modify the emission signal.  A theoretical description
of such a situation can be given in terms of spin-boson type models
\cite{lucke} and will be elaborated elsewhere.

\acknowledgements

We wish to thank J.~Ankerhold, H.~Grabert, C.H.~Mak, R.~Karrlein, 
and G.~Stock for helpful discussions, and acknowledge 
support by the Schwerpunkt ``Zeitabh\"angige Ph\"anomene und Methoden
in Quantensystemen der Physik und Chemie"
of the Deutsche Forschungsgemeinschaft (Bonn).

\appendix
\section*{Variances}

This appendix contains the expressions for the variances
appearing in the general reduced density matrix (\ref{exac}).
Depending on the initial preparation, we get different 
results as described below. 

For a {\sl factorized preparation}, the three independent
expectation values are given by Eqs.~(\ref{qdef}-\ref{p1}).
The various quantities appearing therein read as follows.
For a specific spectral density $J(\omega)$,
the bath correlation function is
\[
L(t) = \frac{1}{\pi} \int_0^\infty d\omega\,
J(\omega) \{ {\rm coth}(\omega\hbar\beta/2) \cos(\omega t)
-i \sin(\omega t) \} \;,
\]
and the Laplace-transformed damping kernel  reads,
\begin{equation}\label{dkern}
\hat{\gamma}(z) = \frac{2z}{\pi m} \int_0^\infty d\omega'\,
\frac{J(\omega') }{\omega'} \frac{1}{\omega^{\prime 2} + z^2} \;.
\end{equation}
The functions $K_q(t)$ and $K_p(t)$ are defined as
\begin{eqnarray}\label{kq}
K_q(t) &=& \int_0^t dt'\, G(t') \int_0^t dt^{\prime\prime} 
\, G(t^{\prime\prime}) L'(t'-t^{\prime\prime} ) \;, \\
\label{kp}
K_p(t) &=& \int_0^t dt'\,\dot{G}(t') \int_0^t dt^{\prime\prime} 
\, \dot{G}(t^{\prime\prime}) L'(t'-t^{\prime\prime} ) \;,
\end{eqnarray}
with $L'(t) =\, {\rm Re} L(t)$ and $G(t)$ being the inverse 
Laplace transform of 
\begin{equation}\label{gz}
\hat{G}(z) = \left[ z^2 + z \hat{\gamma}(z) + \omega_0^2 \right]^{-1} \;.
\end{equation}
In order to ensure that the wave packet 
relaxes to $q=q_0$ at long times, the bath
must fulfill the condition $G(t\to \infty)=0$.
Otherwise it merely leads to a mass renormalization
but not to truly dissipative behavior.

To obtain the  time-dependent variances
(\ref{sigma1}) and (\ref{p1}) in practice, it is necessary
to  find a more convenient form of Eqs.~(\ref{kq}) and (\ref{kp}).
By following Ref.~\cite{grabert88}, we obtain  ($\nu_n=2\pi n/\hbar\beta$)
\begin{eqnarray} \nonumber
K_q(t) &=& \frac{m}{\hbar \beta} \Bigl [
2\sum_{n=0}^\infty \Bigl \{\left(1-
\hat{G}^{-1}(\nu_n) \int_0^t dt' \, G(t') e^{-\nu_n t'} \right)
\\ \nonumber &\times& \int_0^t dt' \, G(t') e^{\nu_n t'} 
- G^2 (t) + \int_0^t dt' \, G(t') e^{-\nu_n t'} \Bigr\} \\
 &-& \Bigl \{ \left(2 - \omega_0^2\int_0^t dt'\,
G(t') \right) \int_0^t  dt'\, G(t') \Bigr\} \Bigr] 
 \label{kq1}
\end{eqnarray}
and 
\begin{eqnarray} \nonumber
K_p(t) &=& \frac{m}{\hbar \beta} \Bigl [
2\sum_{n=0}^\infty \Bigl \{\left(\nu_n -
\hat{G}^{-1}(\nu_n) \int_0^t dt' \, \dot{G}(t') e^{-\nu_n t'} \right)
\\ \nonumber &\times& \int_0^t dt' \, \dot{G}(t') e^{\nu_n t'} 
-\nu_n \int_0^t dt' \, \dot{G}(t') e^{-\nu_n t'} \\
&-& \dot{G}^2 (t) + 1\Bigr\} +\omega_0^2 G^2(t) \Bigr] \;.
 \label{kp1}
\end{eqnarray}
For $t\to \infty$, the equilibrium variances then follow as
\begin{eqnarray} \label{equ_values}
\sigma_\beta &=& \frac{2}{m\beta} \sum_{n=-\infty}^\infty
\hat{G}(|\nu_n|) \;,\\ \label{pp2}
\langle [\Delta p]^2\rangle_\beta &=& \frac{m}{\beta} 
\sum_{n=-\infty}^\infty [1-\nu_n^2 \hat{G}(|\nu_n|)] \;.
\end{eqnarray}

For the {\sl correlated preparation} discussed in Sec.~\ref{corrp},
while $\langle q(t) \rangle$ stays the same as under the factorized
preparation, the variances now read
\begin{eqnarray} 
\sigma (t) &=& \frac{2\hbar}{m} \Bigl( \Lambda_{G} \dot{G}^2(t) + \Omega_{G}
G^2(t) + B(t) \\ & +& 2 \Bigl\{\Lambda_{G}
\dot{G}(t)\, G(t) \, C_{1}^+(t) - G^2(t) \, C_{2}^+(t)\Bigr\}
\nonumber \\ \nonumber &+& \frac{1}{m}
K_{q}(t)\Bigr) \;, \\  
\langle [\Delta p]^2 \rangle (t) & =& \hbar m \Bigl(\Lambda_{G} \ddot{G}^2(t) +
\Omega_{G} \dot{G}^2(t) + S(t) \\ &+&
2\Bigl\{\Lambda_{G} \ddot{G}(t) \bar{C}_{1}(t) - \dot{G}(t) \bar{C}_{2}(t)
\Bigr\} + \frac{1}{m} K_{p}(t)\Bigr) \nonumber \;.
\end{eqnarray}
Herein the various quantities are given as follows \cite{grabert88}, 
\begin{eqnarray*}
\Omega_G &=& \frac{1}{\hbar \beta} \sum_{n=-\infty}^{\infty}
\hat{G}_{G}(|\nu_{n}|)(\omega_{G}^{2} + |\nu_{n}|
\hat{\gamma}(|\nu_{n}|)) \;, \\
\Lambda_G &=& \frac{1}{\hbar \beta} \sum_{n=-\infty}^{\infty}
\hat{G}_{G}(|\nu_{n}|) \;,\\
B(t) &=& \frac{1}{\hbar \beta} \sum_{n=-\infty}^{\infty}
       \hat{G}_{G}(|\nu_{n}|)
       \int_{0}^{t} ds \int_{0}^{t} du \;[g_{n}(s) g_{n}(u) \\
       &-& f_{n}(s)
        f_{n}(u)] G(t-s)G(t-u)  \;, \\
S(t) & =& \frac{1}{\hbar \beta} \sum_{n=-\infty}^{\infty}
       \hat{G}_{G}(|\nu_{n}|)
       \int_{0}^{t} ds \int_{0}^{t} du \;[g_{n}(s) g_{n}(u)
       \\ &-& f_{n}(s) f_{n}(u)] \dot{G}(t-s)\dot{G}(t-u) \;,
\end{eqnarray*}
where $\hat{G}_{G}$ is given by Eq.~(\ref{gz}) with $\omega_{0}$
being replaced by $\omega_{G}$. The functions $g_n$ and $f_n$ are
given by
\begin{eqnarray*}
g_{n}(s) &=& \frac {1}{m \pi} \int_{0}^{\infty} d \omega\, J(\omega) \frac{2
\omega}{\omega^{2} + \nu_{n}^{2}} \,\cos(\omega s)\;,\\
f_{n}(s) &=& \frac {1}{m \pi} \int_{0}^{\infty} d \omega\, J(\omega) \frac{2
\nu_{n}}{\omega^{2} + \nu_{n}^{2}}\, \sin(\omega s)\;.
\end{eqnarray*} 
Furthermore, we have used the abbreviations
\begin{eqnarray*}
C_1(s) & =& \frac{1}{\hbar \beta \Lambda_G} \sum_{n=-\infty}^{\infty}
\hat{G}_{G}(|\nu_n|) g_{n}(s) \;,\\
C_2(s) & = &\frac{1}{\hbar \beta} \sum_{n=-\infty}^{\infty}
\hat{G}_{G}(|\nu_n|) \nu_n f_{n}(s)\;,\\
C_{i}^{+}(t)&  =& \int_{0}^{t} ds \; C_{i} (s) \frac{G(t-s)}{G(t)} \;,\\
\bar{C}_{i}(t) & =& \int_{0}^{t} ds \; C_{i}(s) \dot{G}(t-s)\;.
\end{eqnarray*}

Next we briefly discuss the case of an ohmic bath with
a Drude cutoff, see Eq.~(\ref{drudes}).
The damping kernel then exhibits exponential decay,
$\gamma(t) = \gamma\omega_D \exp[-\omega_D t]$,
with the Laplace transform
\begin{equation}\label{druded}
 \hat{\gamma}(z) = \frac{\gamma\omega_D}{\omega_D+z}  \;.
\end{equation}
Defining $\lambda_i$ for $i=1,2,3$ as the roots of the 
cubic equation
\begin{equation} \label{cubic}
z^3 - \omega_D z^2 + (\gamma \omega_D+\omega_0^2) z
- \omega_0^2 \omega_D = 0 \;,
\end{equation}
one obtains \cite{karrlein}
\begin{equation}\label{drudeg}
\hat{G}(z) = \frac{z+\omega_D}{(z+\lambda_1) (z+\lambda_2)(z+\lambda_3)} 
= \sum_{i=1}^3 \frac{\Lambda_i}{z+\lambda_i} \;,
\end{equation}
where the coefficients $\Lambda_i$ follow as 
\[
\Lambda_i = \frac{\lambda_i (\omega_D-\lambda_i)}
{2\lambda_i^3 - \omega_D (\lambda_i^2 - \omega_0^2)}\;.
\]
From Eq.~(\ref{drudeg}) we arrive at the simple result 
$G(t) = \sum_i \Lambda_i \exp[-\lambda_i t]$.
All  variances can then be evaluated in closed form \cite{lothar}. 
 As the resulting expressions
are very lengthy but can be straightforwardly obtained by following the
above steps, we refrain from quoting them here.

To locate the coherent-to-incoherent transition, we note that the
cubic equation (\ref{cubic}) has either three real 
solutions, or one real and two complex conjugate 
ones. In the latter case, $G(t)$ and therefore $\langle q(t)
\rangle$ will exhibit coherent oscillations.  
The critical value $\gamma_c$ then follows from the condition $D(\bar{\gamma},
\bar{\omega}_D) =0$, where $\bar{\gamma}=\gamma/\omega_0$,
 $\bar{\omega}_D= \omega_D/\omega_0$, and
\begin{eqnarray}\label{det}
D(\bar{\gamma}, \bar{\omega}_D)&=& 
\bar{\gamma}^3 + \bar{\gamma}^2 \left(
\frac{3}{\bar{\omega}_D} - \frac{\bar{\omega}_D}{4} \right) \\
\nonumber &+& \bar{\gamma} \left(\frac{3}{\bar{\omega}_D^2} - 5
\right) + \frac{1}{\bar{\omega}_D^3} + \frac{2}{\bar{\omega}_D} +
\bar{\omega}_D \;.
\end{eqnarray}
For arbitrary $\omega_0$ and $\omega_D$, there is exactly one positive value
$\gamma=\gamma_c$ solving the condition $D=0$.  

\end{document}